\documentstyle[12pt,aaspp4,graphicx]{article}

\def\pgamma{ \hat \gamma}
\def\gf2{(1+\epsilon)}

\def\etal{{\it{et. al. }}}

\begin{document}
\title{Radiative Efficiencies of Continuously Powered Blast Waves}
\author{Ehud Cohen and Tsvi Piran}
\affil{The Racah Institute of Physics, The Hebrew University, Jerusalem 91904, Israel }
\authoremail{udic@nikki.fiz.huji.ac.il}

\begin{abstract}
  We use general arguments to show that a continuously powered
  radiative blast wave can behave self similarly if the energy
  injection and radiation mechanisms are self similar.  In that case,
  the power-law indices of the blast wave evolution are set by only
  one of the two constituent physical mechanisms.  If the luminosity
  of the energy source drops fast enough, the radiation mechanisms set
  the power-law indices, otherwise, they are set by the behavior of
  the  energy source itself. We obtain self similar solutions for
  the Newtonian and the ultra-relativistic limits. Both limits behave
  self similarly if we assume that the central source supplies energy
  in the form of a hot wind, and that the radiative mechanism is the
  semi-radiative mechanism of Cohen, Piran \& Sari (1998).  We
  calculate the instantaneous radiative efficiencies for both limits
  and find that a relativistic blast wave has a higher efficiency than
  a Newtonian one. The instantaneous radiative efficiency depends
  strongly on the hydrodynamics and cannot be approximated by an
  estimate of local microscopic radiative efficiencies, since a
  fraction of the injected energy is deposited in shocked matter.
  These solutions can be used to calculate Gamma Ray Bursts 
  afterglows, for cases in which the energy is not supplied
  instantaneously.
\end{abstract}

\section{Introduction}
Afterglows from Gamma-Ray Bursts (GRB) have been discovered for 13
GRBs since the first detection of the afterglow of GRB970228 ( e.g.
\cite{Zand98}).  The simplest cosmological fireball afterglow model
(\cite{PacRho93,Katz94,MR97,waxman,mes,piran98}) seems to be in a good
general agreement with the observed behavior (\cite{Wiejers_MR97}), or
at least with the power-law decay. However, several works have tried
to investigate more subtle effects which can change the afterglow
characteristics. In particular, \cite{RM98} and \cite{PMR98} have
tried to explain the deviations from an ideal power-law, and the
variety of light curves using a model in which a source emits shells
with different Lorentz factors, which results in a gradual energy
supply to the shell of shocked matter. In their work they have
neglected the thickness of the shell of shocked matter, and did not
calculate its structure.  However, as evident from the classical
treatment of adiabatic blast waves by \cite{bm}, the evolution of a
blast wave  depends strongly on the detailed structure of the matter
inside it. Therefore, it remains to investigate the complete
hydrodynamics and energy budget of a slow power-law decay blast wave.

The dynamics of blast waves with gradual energy supply is also
relevant for the study of compact steep-spectrum objects (CSS), Gigahertz peak
spectrum objects (GPS), and active galactic nuclei (AGN) radio lobes. The
generally accepted model for these objects (see e.g.
\cite{S74,BC89,BDO97}) describes a blast wave continuously powered by
a jet from an AGN. Similar models have also been used for interstellar
bubbles (\cite{C75},\cite{W77}), for plerionic supernova remnants (see
e.g. \cite{Wei83}) and for galactic super-shells (\cite{RM87}).

In this paper we consider blast waves with a gradual energy supply by
a hot wind. Specifically we obtain a new self similar solution for a
continuously powered radiative blast waves.  Even if this situation
may not be directly applicable to astrophysical objects, it is still
important.  Self similar solutions are simple enough that they may be
solved analytically (sometimes), and are easy to grasp.  Furthermore,
it is likely that a generic blast wave will have tendency to approach
self similarity.

Adiabatic self similar blast waves with energy injection have been
treated previously by \cite{bm} in the relativistic regime.  In the
Newtonian regime Castor, McCray and Weaver (1975) and \cite{OM88} have
investigated both radiative and adiabatic blast waves with energy
injection, but have not treated the case of fast decaying sources,
which did not fit into their self similar framework.  In this paper
we show that the inclusion of radiative losses has a qualitative
influence on the solutions.  Using our new solutions we obtain an
accurate result for the fully radiative steady injected blast waves,
which were treated approximately by \cite{bm}. We add new solutions
for a region of the parameter space which has not been treated yet
neither in the Newtonian nor in the ultra-relativistic limits.

Self similarity appears only if the physical situation can be
characterized by a minimal number of dimensional parameters, such that
physical scales can be constructed only from a single  combination of those
parameters. Therefore, by assuming that a blast wave behaves self
similarly, we can deduce the functional form of different physical
processes. In fact, adiabatic blast waves already exhibit self
similar behavior without additional parameters. This requires that
energy injection and radiation processes in self similar blast waves
will not have intrinsic scales.  Specifically, the injection rate must be
a power-law. 

The radiation process should also be self-similar.  In order to
satisfy that demand, we assume that the radiation process results with a
semi-radiative scenario where the cooling mechanisms are fast
comparable to hydrodynamic time scales (``fast cooling''), but only a
fraction of the energy produced by the shock is radiated away. This
would take place, naturally, in any collisionless shock acceleration
(see \cite{CPS98}, hereafter CPS98). For example, in a GRB afterglow the
cooling time scales  are shorter then the hydrodynamic time scale, (see e.g.
\cite{waxman,mes,sari}), and  a fraction of the energy stays with
shocked protons which do not cool.

We describe our model in Sec. \ref{sec:model}. It is composed of a
central source emitting a wind which interacts with a surrounding
medium and creates a blast wave. We proceed in Sec.
\ref{sec:energy_conserve} by calculating the blast wave energy and the
radiated luminosity assuming self-similarity {\it alone}. In Sec.
\ref{sec:newt_ss} and Sec. \ref{sec:rel_ss} we split the discussion to
the Newtonian and ultra-relativistic limits, assuming the self-similar
radiation mechanism of CPS98.  For each case we obtain the
hydrodynamic solution and the radiative efficiencies (analytic
solution for the ultra relativistic limit, and numerical for the
Newtonian limit). Finally we summarize our results in Sec. \ref{sec:conc}.

\section{The Model}
\label{sec:model}
We consider a spherical semi-radiative blast wave, that appears when
energy is released continuously into an ambient medium.
This results in a strong shock
wave that expands supersonically.  We consider the regime where the
influence of the injected and initial mass is negligible, and that the
pressure of the surrounding medium is small compared to the energy
density of the flow. These assumptions are necessary in order to
obtain a self similar solution (see e.g. \cite{ll}, \S 99).

We assume that the source supplies energy in the form of a hot wind
with a negligible mass. This wind pushes away the surrounding medium,
and creates a cavity. This simple model leads to different qualitative
behaviors, depending on the expansion velocity of the blast wave.  If
this velocity is much lower than the
speed of light, (Newtonian limit) the sound velocity of the ejected
wind can be much higher than the expansion velocity. In this case the
pressure inside the cavity settles fast and becomes isobaric. This
isobaric bubble pushes away the ambient matter and creates a shock
wave. (This model, where the injected wind is treated as an isobaric
interior has been widely used in the study of radio lobes in active
galactic nuclei (\cite{S74})).

In the ultra relativistic case the hot wind has different dynamics.
The sound velocity of the wind is of the order of the speed of light,
but it is always comparable to the expansion velocity. In this case we
assume that the wind is so sparse that its' particles inside the cavity
do not interact, and simply move in straight lines, close to the speed
of light, until they reach the edge of the cavity. Near this edge they
undergo a strong shock, presumably a collisionless one, which
compresses them.  Consequently, this compressed fluid produces a shock
which advances into the ISM.

\begin{figure}
\begin{center}
\includegraphics[width=15cm]{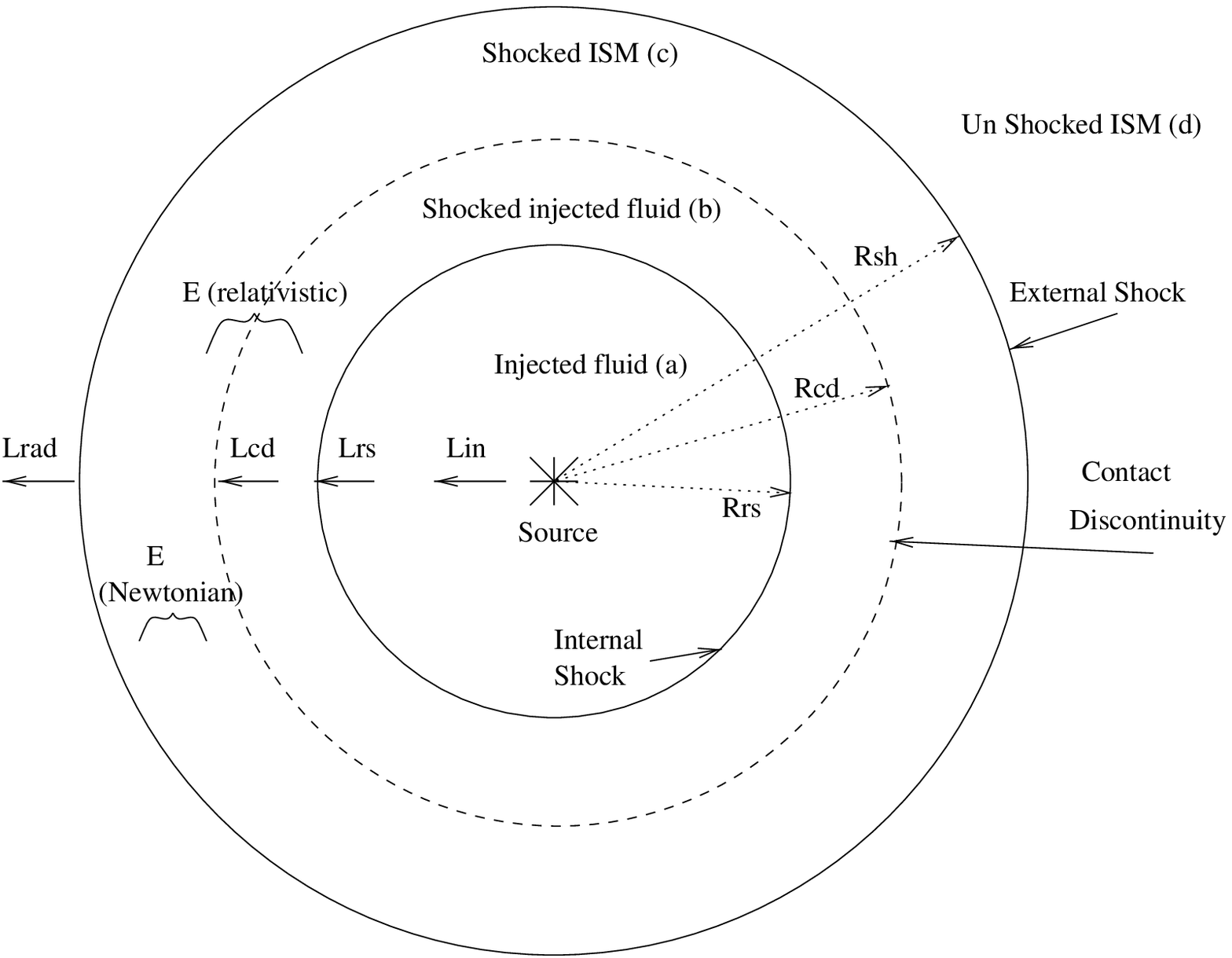}

\caption{
  A schematic drawing of the regions in an energy injected blast wave.
  The regions are: (a) injected hot wind; (b) shocked wind; (c)
  shocked ISM; and (d) ambient ISM. The boundaries between the regions 
  are a reverse shock at $R_{rs}$, a contact discontinuity at
  $R_{cd}$, and a forward shock at $R_{sh}$.
  The central source ejects massless hot wind with a luminosity  of $L_{in}$. Energy is
  transfered at a rate $L_{rs}$ to the shocked matter and at $L_{cd}$ to 
  the shocked ISM.
  The blast wave emits radiation with 
  a luminosity $L_{rad}$ from the shock front.
  In the Newtonian solution $E$ denotes the energy stored in shocked
  ISM, where in the ultra relativistic limit $E$ includes the energy
  stored in shocked wind as well.
  Region (a) is negligible in the Newtonian regime due to the infinite sound velocity
  of the wind and  $L_{rs} = L_{in}$. \label{fig:schematic} }
\end{center}
\end{figure}

Both in the Newtonian and in the ultra relativistic limits the blast
wave is composed of four regions. (a) The injected wind:
contains wind that did not interact, yet, with  the blast wave. In the Newtonian
limit the reverse shock reaches the center, and  this region disappears.
(b) Shocked wind: contains wind that was compressed due to
the interaction with shocked ISM.  It  is separated from region
(a) by a reverse shock. (c) Shocked ISM: contains ISM that has been compressed
by the forward shock at the blast wave front. A contact discontinuity
separates this region from the shocked wind region. (d) ISM at rest. A 
schematic view of the four regions appears in Fig. \ref{fig:schematic}.

\section{Self similarity in energy injection cases}
\label{sec:energy_conserve}
We look for self similar solutions for semi-radiative blast waves with
central energy supply.  If a blast wave contains only one of these two
mechanisms ( energy injection and radiation ), each of them leads to
a different temporal behavior. As self similarity does not allow the
existence of several time scales in the solution, it is apriori not
known whether such solutions exist.

Self similarity requires a dimensionless energy injection mechanism.
We therefore deduce that the energy is supplied to any interval
along the self similar solution 
at a rate $L \propto t^{\sigma}$.  The energy transfer rates along the
self similar hydrodynamic profile ( especially $L_{rs}$ and $L_{cd}$ )
scale the same, and differ only by a constant factor, which depends
on the exact solution.  However, this scaling can differ from
$L_{in}$, the energy supplied by the central source, as a fraction of
the injected energy resides in the interior regions.  Region (a) contains
a freely expanding wind. If it exists, it does not depend on the blast
wave parameters. For self similarity the energy transfer rate from
region (a) to region (b-c) must satisfy the self similarity scaling
laws.  However, it is possible that the energy in region (a) will
evolve differently.

Assuming that a self similar solution exists, we can write the energy
loss rate as  
\begin{equation}
\label{Lrad_basic}
L_{rad}=-\kappa {E \over t}, \quad \kappa > 0,
\end{equation}
where $E$ is the energy stored in a fixed interval of the self similar
profile. In the Newtonian limit we choose $E$ to denote the energy
stored in shocked ISM alone (and the corresponding injection rate is
$L_{cd}$) and in the Ultra-relativistic case $E$ denotes the
energy stored in the entire shocked matter (and the corresponding
injection rate is $L_{rs}$).  The value of the constant $\kappa$
depends on this definition, but it is always dimensionless. This
proportionality is evident from self similarity, as there exists only
one way to construct the dimensionality of $L_{rad}$ (energy/time)
using the characteristic parameters. Note that this argument holds for
{\it any} self similar radiation mechanism.

The energy conservation equation includes losses as well as 
injection from the blast wave center,
\begin{equation}
\label{dE_dt}
{dE \over dt} =    L ({ t \over t_L})^{\sigma}     
- \kappa {E \over t}.
\end{equation}
For $\sigma \neq  - 1 -\kappa$  Eq. \ref{dE_dt} has an analytic solution
\begin{equation}
\label{E_of_t}
E = {L  \over \kappa + \sigma + 1} ({t\over t_L})^{\sigma} t + A t^{-\kappa},  
\end{equation}
where $A$ is set by the initial conditions.

We proceed by investigating the asymptotic behavior of this equation
for different values of $\sigma$ and $\kappa$. In cases where the
equations behaves self similarly we find the power-law index of the
energy temporal behavior $E\propto t^{\lambda}$, and obtain the
relations between the various power-law indices. A summary of the
different limits appears in table 1.

For $L=0$, we reproduce the instantaneous energy injection case and
Eq. \ref{E_of_t} trivially becomes a power-law with $\lambda =
-\kappa$. Even in this simple case, an extrapolation toward $t \to 0$
results in an infinite energy. From Eq. \ref{Lrad_basic} it is obvious
that the radiated energy also diverges in this limit. These two
infinities cancel each other, and at finite times the energy is
always finite.
A specific  blast wave with an arbitrary initial energy
evolves to this self similar behavior at late times. Clearly  the
limit $t \to 0$ has no physical relevance and only the late time
behavior is interesting.
It is therefore better to treat the solution from $t=\infty$
backward in time, as this is the region where we expect the blast wave
to evolve according to the self similar solution. At any finite
time all the energies are finite, and  we encounter no infinities using this
view. We discuss $E_{rad}^{\infty}(t)$, which is
the energy that will be radiated from time $t$ to infinity. It is
evident that $E_{rad}^{\infty}(t)=E(t)$, as the energy stored in the
shell decreases to zero when $t \to \infty$.

Sources which scales with $\sigma<-1$ were not treated so far. In this
scenario the injected and radiated energies from $t=0$ are infinite,
similarly to the $L=0$ case. This means that the solution is not
physical near $t=0$. 
As in the $L=0$ case, we overcome the
infinities by discussing $E_{inj}^{\infty}(t)$. The solutions are
divided into three sub-classes, depending on the relation between
$\sigma$ and $\kappa$: 
\begin{itemize}
\item 
If $\sigma > -1-\kappa$ the injected energy decreases slower
than the rate of an un-injected blast wave, and the power-laws are
set by the injection law \footnote{
Even though the power-law index is set by the
injection mechanism and not by the radiative mechanism, the radiated
energy is not negligible. }.
The energies injected and radiated from $t=0$ are infinite. However,
the energy that will be radiated until infinity is always larger than
the stored energy, and the energy which will be injected is larger
(smaller) than the stored energy if $\sigma$ is larger (smaller) than
$-1-\kappa/2$. Note that even if the stored energy is larger than the
energy injected until infinity, the evolution is set by the injected
energy power-law. In both cases $\lambda=\sigma+1$.

\item If $\sigma = -1-\kappa$ then Eq. \ref{E_of_t} is not valid. 
The solution of  Eq.  \ref{dE_dt} for this case is
\begin{equation}
\label{E_of_t2}
E_{shell}=\left({t \over t_L}\right)^{-\kappa}\left[{A+L t_L \log (t/t_L)}\right].
\end{equation}
This equation has no power-law asymptotics, and there is no self
similar solution in this case.

\item In cases where  $\sigma <
-1-\kappa$, the late time behavior is identical to the
instantaneous injection case, and the asymptotic of Eq. \ref{E_of_t} is
a power-law with $\lambda=-\kappa$. Note that in all the sub-cases
with $\sigma<-1$ the stored energy drops to zero with time.
\end{itemize}

Sources  with $\sigma>-1$ were treated by \cite{OM88}.
In this case  $E_{inj}^0(t)$, the energy injected from $t=0$, is finite for
every $t$, but it diverges as $t \to \infty$. This divergence allow us to
neglect any initial energy in the blast wave.  Eq. \ref{E_of_t} again
has a power-law asymptotic, now with $\lambda = \sigma+1$. The energy
stored in the shocked matter is a constant fraction of the energy
injected so far, and the same holds for the radiated energy.  The
energy injected is always larger then the energy stored in the shocked
matter, and the radiated energy from $t=0$ is larger (smaller) than
the stored energy depending if $\kappa$ is larger (smaller) than
$\sigma+1$.  

If the energy is supplied with $\sigma=-1$, the integral of injected
energy is logarithmic, and it diverges in both limits of $t$. However,
the asymptotic of Eq.  \ref{E_of_t} is again a power-law, now with
$\lambda=0$, i.e. the shell holds a constant energy which is set by
initial conditions.

The study of the energy conservation equations revealed the temporal
behavior of the energy if the solution is self-similar.  However,
$\kappa$ cannot be determined by self-similarity arguments alone.  It
is generally determined only by solving the complete hydrodynamic
equations.  We split the discussion of this solution to the Newtonian
and the ultra-relativistic limits.

\section{The Newtonian solution}
\label{sec:newt_ss}
To solve the hydrodynamic equations we use our model (Sec.
\ref{sec:model}), where the energy from the central source is supplied
by an internal fluid with a high temperature and a low density, in the
limit of infinite sound velocity.  The large sound speed of the
injected matter prevents the existence of an interior shock, and the
flow contains only three regions (see Fig.  \ref{fig:schematic}): (b)
an isobaric interior composed of shocked wind; (c) shocked
interstellar medium; and (d) ambient ISM.  The hot internal fluid
pushes the shocked matter at the contact discontinuity and accelerates
it. The shocked matter accretes mass from the ambient medium, and
heats it at the shock front. The heated matter emits a fraction of its
internal energy near the shock front.

If the evolution is governed by the injection mechanism, the blast
wave is characterized by its instantaneous radiative efficiency, which
is the ratio between emitted energy (at the shock front) and the
energy released at the center, {\it at the same time}. Due to the
structure of the blast wave (the three regions), it is helpful to
divide the energy transfer into two stages: (i) energy transfer from the
wind to the shocked matter; and (ii) energy transfer from the shocked
matter to radiation. 

Both stages show instantaneous efficiencies which are larger than
unity if the source luminosity drops fast enough. This is caused by
the delay between the time energy is supplied to the shocked matter
and the time this energy is radiated. For example, if the source shuts
down spontaneously it emits no energy, but energy is still radiated
from the hot shocked matter. This results in an infinite instantaneous
efficiency.

During the expansion, the hot internal fluid (region (b)) performs work on
the shocked matter (region (c)), $L_{cd}=4 \pi r_{cd}^2 p_{cd} v_{cd}$, where $r_{cd},
p_{cd},$ and $v_{cd}$ are respectively the radius, pressure and velocity of
the contact discontinuity. Meanwhile, the internal fluid 
absorbs energy from  the internal source which emits energy
at a rate 
\begin{equation}
\label{non_adiabatic_L}
L_{in}= 4 \pi /3 [{d( r_{cd}^3 p_{cd} / ( \pgamma - 1)) \over dt} + p_{cd} {d r_{cd}^3
  \over dt} ].     
\end{equation}
Prior to solving the self similar hydrodynamics, we use the self
similar relations  to calculate the efficiency $L_{cd}/L_{in}$.  
A Newtonian blast wave must
evolve according to the self similar variable (see e.g. \cite{ll}, \S 106):
\begin{equation}
\label{n_ss}
\xi = r \left[ { \rho_1 \over t^{2+\lambda} (E_0 / t_0^{\lambda})}
\right ]^{1/5} = \xi_0 r/R_{sh},
\end{equation}  
where $\rho_1$ is the density of the surrounding medium, and $R_{sh}$
is the radius of the shell. The self similar quantities, which turn
the hydrodynamic equations into coupled ordinary differential
equations (ODEs) (see e.g. \cite{ll}, \S 106) are
\begin{eqnarray}
\label{n_ss_subs}
\rho(r,t) &=&  \left( { \pgamma + 1 \over \pgamma - 1 } \right) \rho_1
\alpha(\xi), \nonumber \\
u(r,t) &=&    { 2 \over \pgamma + 1 }  U_{sh}  {r \over R_{sh}(t)} v(\xi),\\
p(r,t) &=&   { 2 \over \pgamma+1 } \rho_1 U_{sh}^2  ({r \over
  R_{sh}(t)})^2 p(\xi) \nonumber.
\end{eqnarray}
Keeping in mind that the ratio $r_{cd} / R_{sh}$ is constant (again
due to self similarity), we substitute $p(r,t)$ into Eq.
\ref{non_adiabatic_L} and  use Eq. \ref{n_ss} to obtain
\begin{equation}
\eta_{in} = L_{cd} / L_{in} = { 3 (\pgamma-1)(\lambda+2) \over 2 (\lambda-3) + 3 \pgamma (\lambda+2)}.
\end{equation}
For $\lambda > 0$, (which corresponds to energy injection cases with
$\sigma > -1$), $\eta_{in}<1$, as expected.  However,
for $\lambda <0$, $\eta_{in}>1$, as the radiated energy is mainly
supported by energy that is already in the shocked matter.
For $\lambda = -6 ( \pgamma-1)/(2+3 \pgamma)$, $\eta_{in}$ is
infinite. This corresponds to the \cite{OM88} solution of
adiabatic interior with no internal energy supply which pushes a fully
radiative shell.

This efficiency reveals the ratio between the injected energy $L_{in}$
and $L_{cd}$, the rate in
which energy is supplied to the shocked ISM. To obtain the
instantaneous global efficiency we use Eq. \ref{E_of_t}, and find that
$\eta_{rad} = {L_{rad}/ L_{cd}} = \kappa / (\kappa+\sigma+1)$.

To obtain  $\kappa$ we need
the full
hydrodynamic solution.  We use the energy loss rate of the
semi-radiative model (CPS98)
\begin{equation}
\label{n_rad}
L_{rad} = -2 \pi \rho_1 U_{sh}^3 R_{sh}^2 \epsilon,
\end{equation}
substitute the self similar relations of Eq. \ref{n_ss}, and obtain
\begin{equation}
\label{n_rad_ss}
\kappa= - 2 \pi ({2+\lambda \over 5})^3 \xi_0^5 \epsilon.
\end{equation}
The instantaneous efficiency is than 
\begin{equation}
\eta_{tot}={L_{rad} \over L_{in}} =  {
3 (\pgamma-1)(\lambda+2) / [ 2 (\lambda-3) + 3 \pgamma (\lambda+2)]
  \over 1 + 125 (1+\sigma)/[ (3+\sigma)^3  2 \pi \xi_0^5 \epsilon  ]},
\end{equation}
where $\xi_0$ is  the self-similar position of the shock.

To obtain $\xi_0$ and the instantaneous efficiency, we 
solve the self-similar equations. The solution is obtained by
integrating numerically the self similar hydrodynamic ODEs from the
shock down to the contact discontinuity (where the fluid does not move
on the self similar profile). The boundary conditions for this
integration are the semi-radiative shock conditions (CPS98)
\begin{equation}
\label{n_ss_initial}
\alpha(\xi_0)={1 \over  1 -  \delta}, \quad v(\xi_0) =  1 + {
  \pgamma-1 \over 2} \delta ,
\quad p(\xi_0) = 1 + { \pgamma-1 \over 2} \delta,
\end{equation}
where $\delta$ is related to the radiative efficiency by
\begin{equation}
\label{n_epsilon}
\epsilon(\delta) = { \delta \over 1+\pgamma }\left[{2+
    (\pgamma-1)\delta}\right].
\end{equation}
We find $\xi_0$ using the  normalization condition which
equates the energy defined in the self-similar parameter
(Eq. \ref{n_ss}) to the energy stored in the shocked matter
\begin{equation}
\label{n_E_tot}
E(t)= \rho_1 R_{sh}^3 U_{sh}^2 { 8 \pi \over \xi_o^5 (\pgamma^2-1)}
\int_{\xi_{cd}}^{\xi_0} (  p(\xi)  + { \alpha(\xi) v(\xi)^2 } )  \xi^4
d\xi.
\end{equation}
If the evolution is governed by the radiation mechanism,  
then $\lambda$ is not known prior to solving the hydrodynamics. In that case 
in addition to Eq. \ref{n_E_tot} we use $\dot E = -L_{rad}$, and obtain both 
$\lambda$ and $\xi_0$ simultaneously.

\begin{figure}

\begin{center}
\includegraphics[width=15cm]{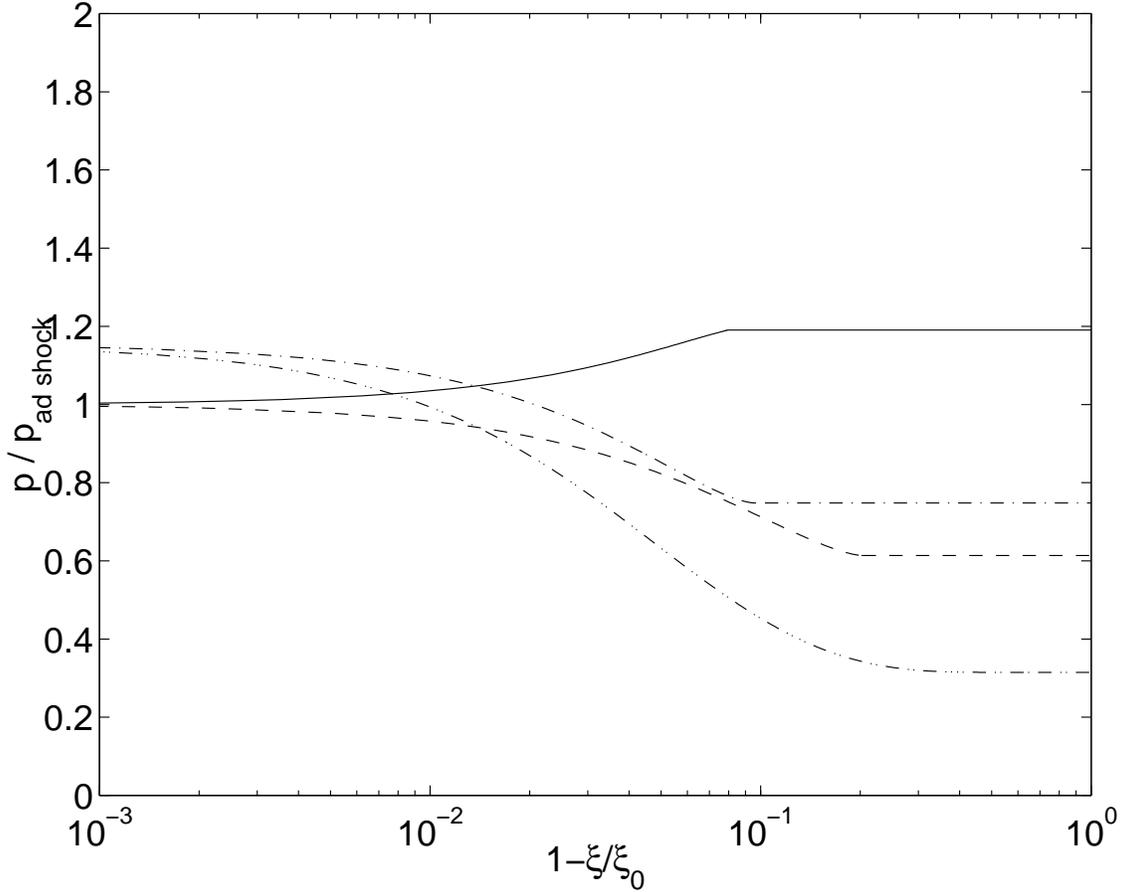}
\caption{
  The pressure $p$ behind the shock front ($\xi=\xi_0$) normalized to
  the pressure near the shock in an adiabatic blast wave, for four
  cases: (i) energy injection $L \propto t^3$, adiabatic (solid); (ii)
  energy injection $L \propto t^{-1/2}$, adiabatic (dashed); (iii)
  energy injection $L \propto t^{-1/2}$, semi-radiative $\epsilon=0.4$
  (dashed-dotted); (iv) energy injection $L \propto t^{-5/4}$
  semi-radiative $\epsilon=0.4$ (dashed double-dotted).  The blast
  wave center $(\xi=0)$ appears on the right side of the graph, where as
  the front shock ($\xi=\xi_0$) is approached on the left.  The
  isobaric region contains the shocked wind, where the rest of the
  blast wave contains shocked ISM.\label{fig:newt_prof1} }
\end{center}
\end{figure}

\begin{figure}

\begin{center}
\includegraphics[width=15cm]{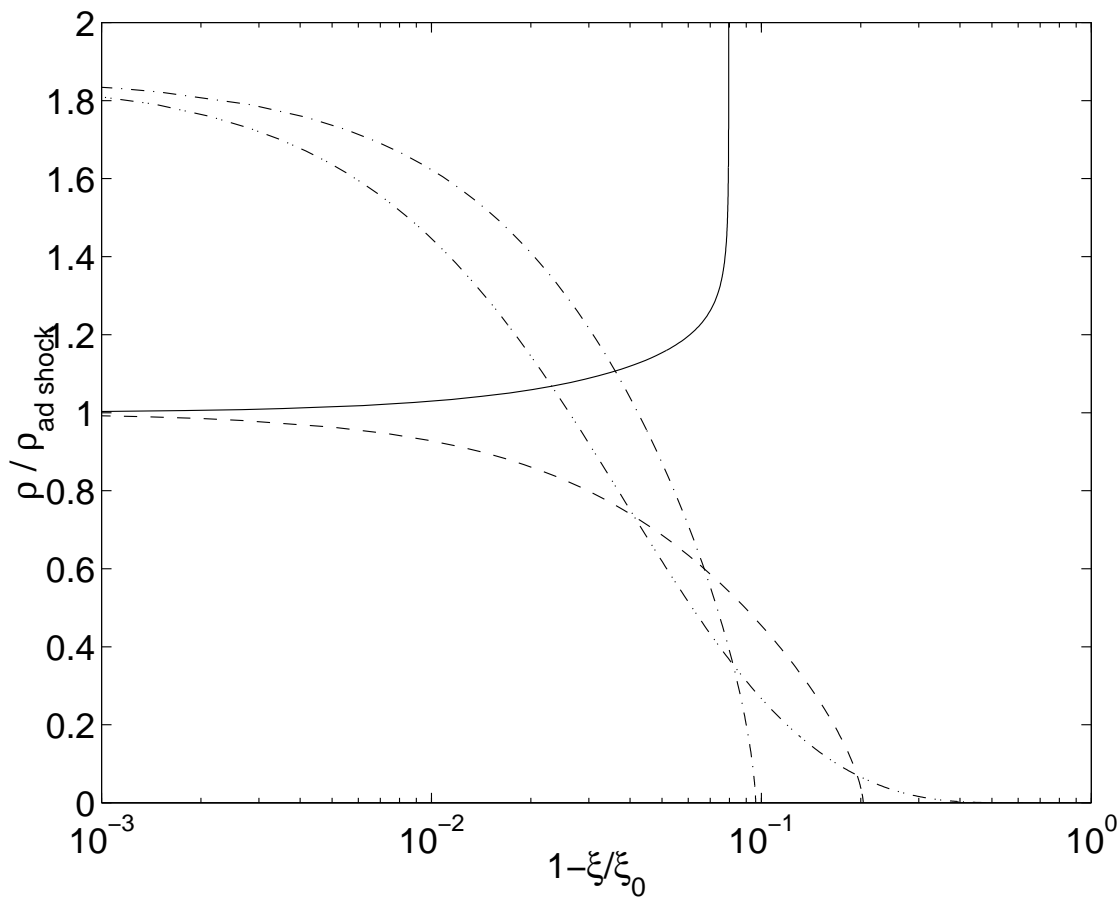}
\caption{
  The normalized density $\rho$ behind the shock front ($\xi=\xi_0$)
  for the cases shown in Fig. \ref{fig:newt_prof1}.
  Cases with density that drops to zero correspond to
  $\sigma<3$. At higher injection rates the density diverges to
  infinity at the contact discontinuity. The
  density at the shock front ($ \xi=\xi_0$ ) is set by the
  semi-radiative conditions, and it does not depend on the injection
  mechanism. \label{fig:newt_prof2} }
\end{center}
\end{figure}

\begin{figure}

\begin{center}
\includegraphics[width=15cm]{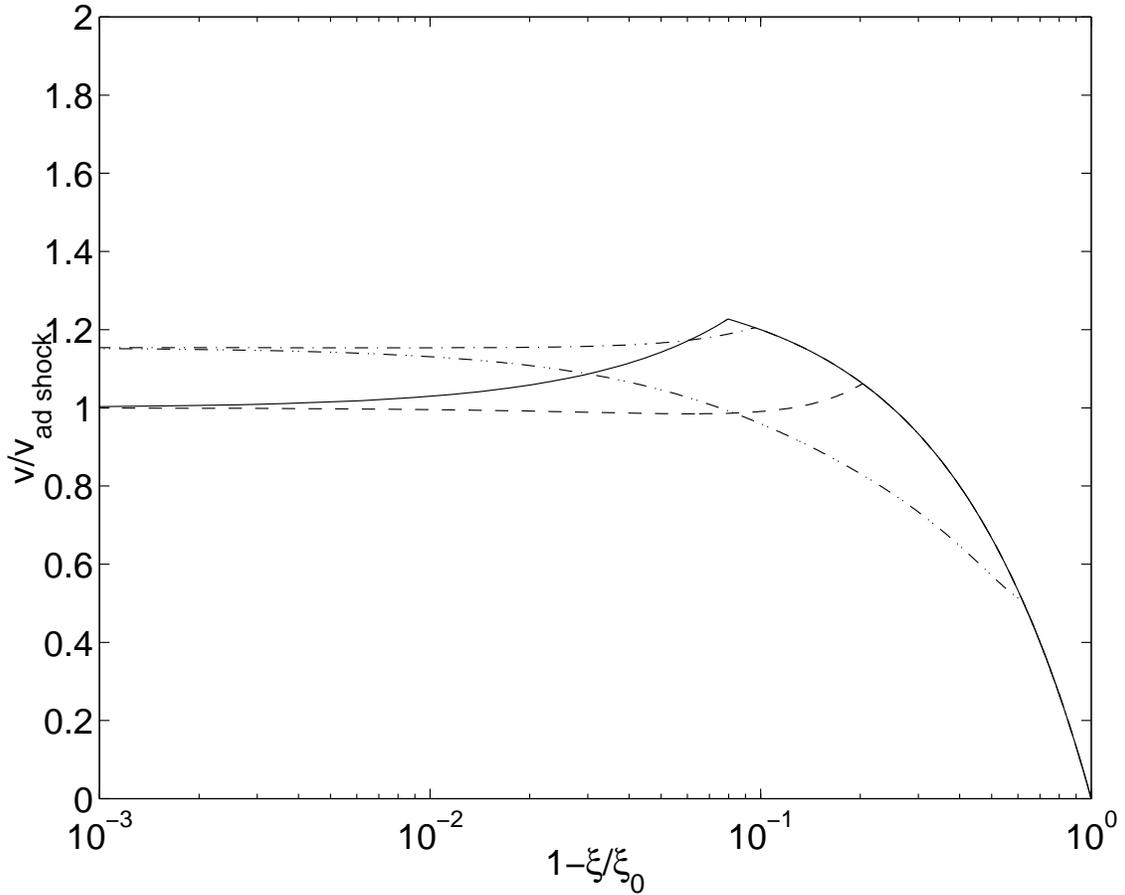}
\caption{
  The normalized velocity $v$ behind the shock front ($\xi=\xi_0$)
  for the cases shown in 
  Fig. \ref{fig:newt_prof1}. 
  All the graphs coincide on the curve $v=U_{sh} \ \xi/\xi_0$, which is
  relevant both for the interior hot bubble and for the contact discontinuity. 
\label{fig:newt_prof3}}
\end{center}
\end{figure}

We present the hydrodynamic solutions for several cases in Fig.
\ref{fig:newt_prof1}-\ref{fig:newt_prof3} and in Table \ref{table:newt}.  In all cases
where the evolution is set by the energy injection mechanism, a contact
discontinuity appears in the solution.  If the injected energy
decreases fast enough with time, the shocked ISM profile is similar to
the one found in instantaneously injected  blast waves.  The temperature rises
monotonously toward the center, and it diverges at the contact
discontinuity.  
The velocity is monotonous, and the fluid near the shock front has the 
maximal velocity.
At the contact
discontinuity the fluid is at rest relative to the profile. The
density  is also monotonous, and it reaches the contact
discontinuity with a zero gradient. This results in a smooth transition
between the shocked wind (which composes the internal isobaric
bubble), and the shocked ISM.  Effectively, there is no discontinuity
at the contact discontinuity.

The boundary between the shocked ISM and the internal hot isobaric
bubble depends on the energy injection rate.  As the injected energy
becomes more dominant, i.e., it is injected with higher $\sigma$, the
matter tends to concentrate in a narrower shell near the front shock. If
$\sigma < { (8-9 \pgamma) /( 4+3 \pgamma)}$, the density gradient
still vanishes at the contact discontinuity, which contains no jumps.
For $\sigma$ larger than this value, the density gradient is infinite,
and the shocked matter region ends abruptly at the contact
discontinuity. If $\sigma>2$, the density it self also diverges to
infinity. Despite these infinities the total mass and energy stored in
the self similar solution are finite, and the solution is a proper
one.  Increasing $\sigma$ also changes the velocity profile
considerably, as for large enough $\sigma$ the maximal fluid velocity
is not at the shock front, but at the contact discontinuity.

Inclusion of a radiative mechanism decreases the shell width, with the
limit of zero width if $\epsilon \to 1$.  The qualitative effect of
radiation is mainly observed in cases where $\sigma<-1$. There, 
the solution is identical to the evolution of a blast wave with
instantaneous energy release, unless the radiative efficiency is high
enough. If $\kappa>1+\sigma$, the evolution is set by the injection
mechanism, and the blast wave has the same structure as ones with $\sigma>-1$.

\begin{figure}

\begin{center}
\includegraphics[width=15cm]{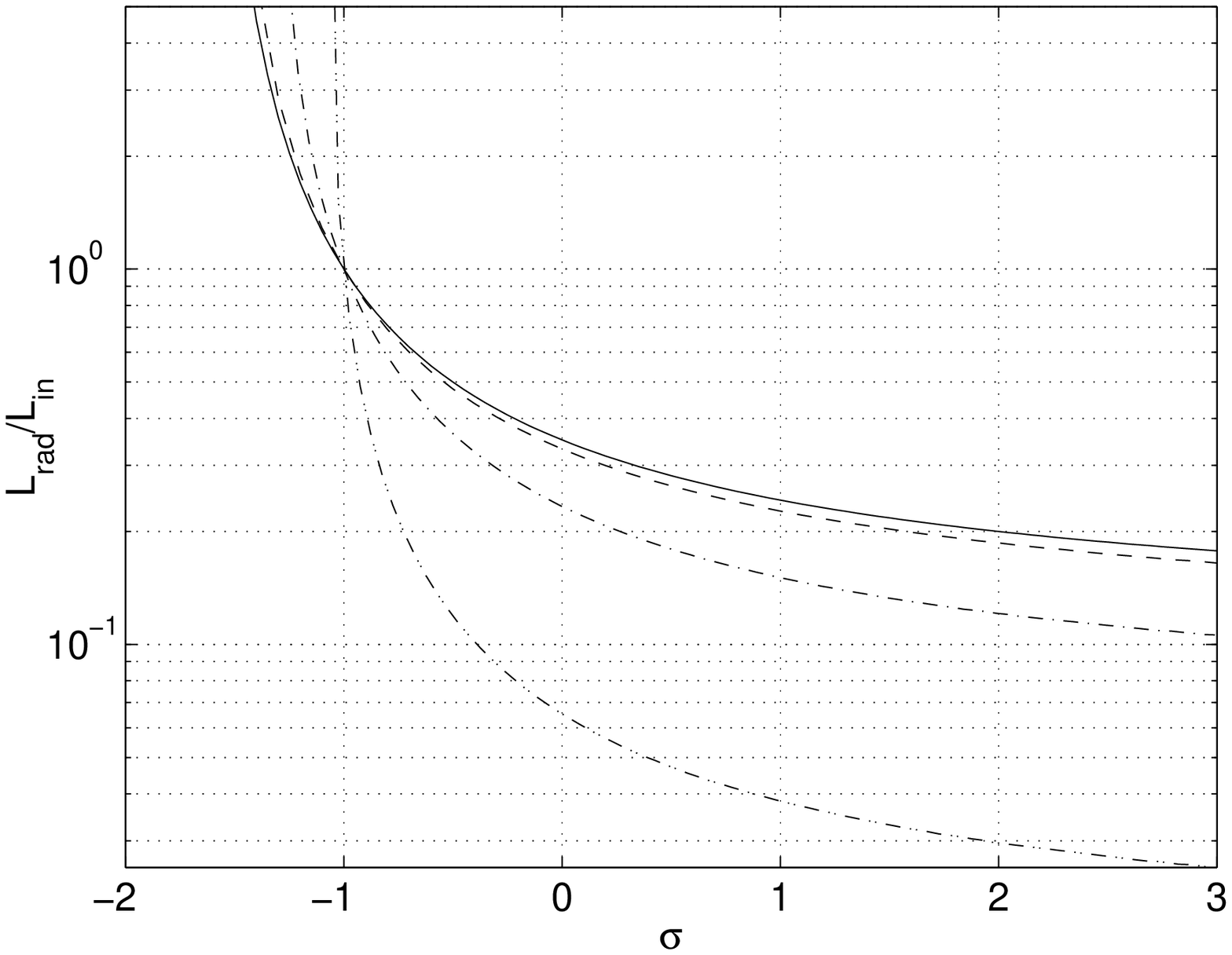}
\caption{
  The instantaneous radiative efficiency vs. energy injection rate of a
  Newtonian blast wave, for various semi-radiative efficiencies: (i)
  $\epsilon=0.1$ (dashed double-dotted); (ii) $\epsilon=0.5$ (
  dashed-dotted); (iii) $\epsilon=0.9$ ( dashed); (iv) $\epsilon=1$
  (solid).\label{fig:newt_eff}  }
\end{center}
\end{figure}

The instantaneous efficiency as a function of $\sigma$ is depicted in
Fig.  \ref{fig:newt_eff}. The efficiency is larger than unity if
$\sigma<-1$ as the main source for radiation is the hot shocked matter
which was heated at earlier times, and not the energy which is
currently being supplied.  If $\sigma>-1$, the main contribution to the
shocked matter energy is the recently released wind. The radiated
energy is therefore directly connected to the wind released at that
time, and the efficiency is lower than unity. 
It appears that the hydrodynamics are an important sink for injected
energy. For example, a steady source with a semi radiative mechanism
of $\epsilon=0.5$, results in an overall efficiency of $0.34$ if $\pgamma=5/3$,
and $0.19$ is $\pgamma=4/3$.

The limit of $\epsilon=1$ cannot be calculated numerically because of
the divergence in the semi-radiative boundary conditions. However, 
a complete  analytic solution exists for this case (see  appendix),
assuming that the whole ISM is concentrated in a narrow shell. The
limit of the numeric solution with $\epsilon \to 1$ reaches this 
solution.

\section{The Ultra-Relativistic solution}
\label{sec:rel_ss}
If the blast wave expansion velocity is ultra-relativistic, the sound speed
of the wind cannot be considered infinite, an internal
shock exists, and the blast wave exhibits the four regions discussed
in Sec. \ref{sec:model}.  The inner region of the blast wave contains
wind which flows unaltered inside the cavity until it reaches the blast
wave edge. This wind has no dependence upon the blast wave parameters,
and it can therefore scale differently in time.

In the derivation of the hydrodynamic solution we follow the footsteps
of \cite{bm}. 
We assume that the inner source varies with time as
$L_{in}=L_0 t_e^q$, and look for a self similar solution for the shocked
matter (wind and interstellar medium) {\it only}.  
{  According to our model, the shocked matter flows
  adiabatically in the entire self similar region of the blast wave}.
The boundary conditions
for the  solution are found using the internal wind, as
the energy supplied to the shell by the wind is equal to the sum of
the radiated energy and the energy stored in the shell. Boundary
conditions for the forward shock are found using the semi-radiative
jump conditions (CPS98).

Self similar solutions for relativistic blast waves were found
(\cite{bm}) to scale according to the self-similar parameter
\begin{equation}
  \label{r_chi_def}
  \chi = [ 1+2(m+1)\Gamma^2(t)](1-r/t), \quad m>-1
\end{equation}
where $\Gamma$ is the Lorenz factor of the shock, which scales as
$\Gamma^2 \propto t^{-m}$.  A blast wave which evolves with $m<-1$ has
no time-like relation between the central source and the blast wave
edge, and thus cannot be reached by injecting energy from a central
source.  However, in other cases, especially with a steep ambient
density gradient, a faster acceleration is possible.  This type of
solutions will be discussed in Cohen, Piran \& Sari (1999).

We start by identifying the different regions in the self similar
solution.  The reverse shock, which separates the self similar portion
of the blast wave from the inner wind must evolve self similarly, and
occur at a fixed self similar location $\chi=\chi_{rs}$. The velocity
of the reverse shock is therefore $v=dr(\chi_{rs},t) / dt$, and its
Lorentz factor is $\Gamma_{rs} =\Gamma / \sqrt{\chi_{rs}}$.  Moreover,
due to the high velocity of the wind, the reverse shock is strong and
the shocked fluid must leave the shock with a velocity of $c/3$
(\cite{ll}, \S 135), which translates into a Lorenz factor of
$\gamma=\sqrt{2} \Gamma_{rs}$ in the unshocked fluid frame.

We can therefore use the self similar substitutions for the pressure
($p$), Lorenz factor ($\gamma$), and density ($\rho$) (BM76),
\begin{eqnarray}
\label{r_ss_subs}
p(r,t) &=& {2 \over 3} \rho_1 \Gamma(t)^2 f(\chi), \nonumber \\ 
\gamma(r,t)^2  &=& {1 \over 2}  \Gamma(t)^2  g(\chi), \\ 
\rho(r,t) \gamma(r,t) &=& 2 \rho_1 \Gamma(t)^2 h(\chi) \nonumber
\end{eqnarray}
to obtain a condition for the reverse shock, $g(\chi_{rs})
\chi_{rs}=4$.  Similarly, for the contact discontinuity, where the
fluid is at rest relative to the self similar profile, we obtain
$g(\chi_{rs}) \chi_{rs}=2$.  These relations enable us to identify the
composition of a certain self similar interval (shocked ISM or shocked
wind) simply by  the term $g(\chi) \chi$.

The wind, which advances with a Lorenz factor which is much larger
than $\Gamma_s$, overtakes the shocked shell with a velocity
difference of $1/2\Gamma_s^2$. The rate in which energy is supplied to
the shell is therefore $dE/dt = L_{in}(t_e)/2\Gamma_s^2$, where $t_e$
accounts for the delay between the emission of the energy, and the
impact at the reverse shock.

We start by obtaining an equation of motion using the energy
conservation equation. The energy stored in shocked matter is
\begin{equation}
\label{rel_ener}
E = \int_0^{R(t)} { 4 \pi \over 3}  r^2  e (4\gamma^2(r,t)-1)  dr=
 8 \pi \rho_1 \alpha_s \Gamma^2 t^3 / 3 (m+1),
\end{equation}
where we define the dimensionless energy integral
$$
\alpha_s \equiv \int_1^{\chi_s}fg d\chi.
$$
Using this energy and the semi radiative model we write the energy
conservation equation as
\begin{equation}
\label{rel_ener_cons}
dE/dt = L(t_e)/2\Gamma_s^2 - 8 \pi \rho_1 R_{sh}^2 \Gamma^2 \epsilon /3,
\end{equation}
and obtain
\begin{equation}
\label{G_of_t}
\Gamma^2 = K ({3 L_0 \over 16 \pi \rho})^{1/(q+2)} t^{(q-2)/(q+2)},
\end{equation}
where
$$
K=[ 2^{3q} (\alpha_s(q+1)+\epsilon) / (q+2)^q ]^{-1/(q+2)} \chi_s^{(q+1)/(q+2)}.
$$

To find $\alpha_s$ we solve the hydrodynamic self similar ODEs (The
analytic solution is given in appendix B.). The boundary conditions at
the shock front ($\chi=1$), are the semi-radiative boundary conditions
(CPS98)
\begin{equation}
f(1)=1, \quad g(1)=1+\epsilon, \quad h(1)={1+\epsilon \over 1-\epsilon}.
\end{equation}
The hydrodynamic profile begins at the shock with these boundary
conditions, and $g(\chi) \chi = 1+\epsilon$. Near the shock the fluid
flows slower than the self similar profile.  Toward the blast wave
center the fluid decelerates relative to that profile, where at
$g(\chi) \chi=2$ the fluid is at rest. The contact discontinuity
resides in this location.  Closer to the center, the fluid accelerates
(the fluid here moves faster then the self similar profile), until it
reaches the reverse shock where $g(\chi_{rs}) \chi_{rs}=4$.

We present the hydrodynamic solutions for several cases in Fig.
\ref{fig:rel_prof1}-\ref{fig:rel_prof3} and in table \ref{table:rel}.
The solutions show the same qualitative behavior as the Newtonian
blast waves.  If the evolution is set by the injection mechanism, the
contact discontinuity and the reverse shock exist. However, if the
injected energy decreases fast enough with time, the shocked ISM
profile is similar to the one found in blast waves without a
continuous power supply. The density, pressure and velocity are
monotonous, and reach the maximum at the shock front. The density
vanishes at the contact discontinuity, with a gradient which diverges
to $-\infty$.  The same structure appears also for $q<-1$, as long as
the evolution is set by the injection mechanism.  If $q>2$, the
density itself diverges to infinity at the contact discontinuity.
High injection rate also results in a velocity profile with a maximum
at the contact discontinuity.

An immediate consequence of the solution applies for the steady
injection fully radiative limit.  \cite{bm} (Eq. 82 there) have
treated this case assuming that the entire radiation is emitted from
some average location within this layer.  We take into account the
momentum losses due to radiation through the entire cooling layer, and
obtain
\begin{equation}
\Gamma^4 =  0.79 L_{in} / 8 \pi \rho_1 R^2.
\end{equation}

Similarly to the Newtonian case, we would like to characterize a blast
wave by its radiative efficiency.  Using Eq. \ref{rel_ener} and Eq.
\ref{rel_ener_cons} we calculate the ratio between the radiated energy
rate and the energy supply rate to the shocked matter (which includes
shocked ISM and shocked internal fluid), and obtain
\begin{equation}
\label{rel_eta_rad}
\eta_{rad} = {L_{rad} \over L_{rs}}={1 \over 1 + \alpha_{s} (1+q) / \epsilon}. 
\end{equation}
Note that prior to the substitution of $\alpha_s$ from the full hydrodynamic solution,
we can use  a zero order approximation and assume that  $\alpha_s$ is constant.
We can then immediately deduce that similarly to the Newtonian case, the efficiency 
rises with increasing $\epsilon$ and decreasing $q$.

The instantaneous efficiency is intrinsically ill-defined in the ultra
relativistic case.  The luminosity scales with time as
$t^{(2+3q)/(2+q)}$ (using Eq. \ref{G_of_t}), which is generally
different from the source behavior $t^q$.  Therefore, the overall
radiation efficiency is not constant in time and cannot be used to
characterize the blast wave.  However, due to the relativistic motion
of the shell, the radiation it emits is measured by the observer at a
different time, $t_{obs}=t/2/(m+1)/\Gamma^2$. Still, the central
source does not move. If it had been measured directly by an
observer, it would evolve with $t$, and not $t_{obs}$.

\begin{figure}

\begin{center}
\includegraphics[width=15cm]{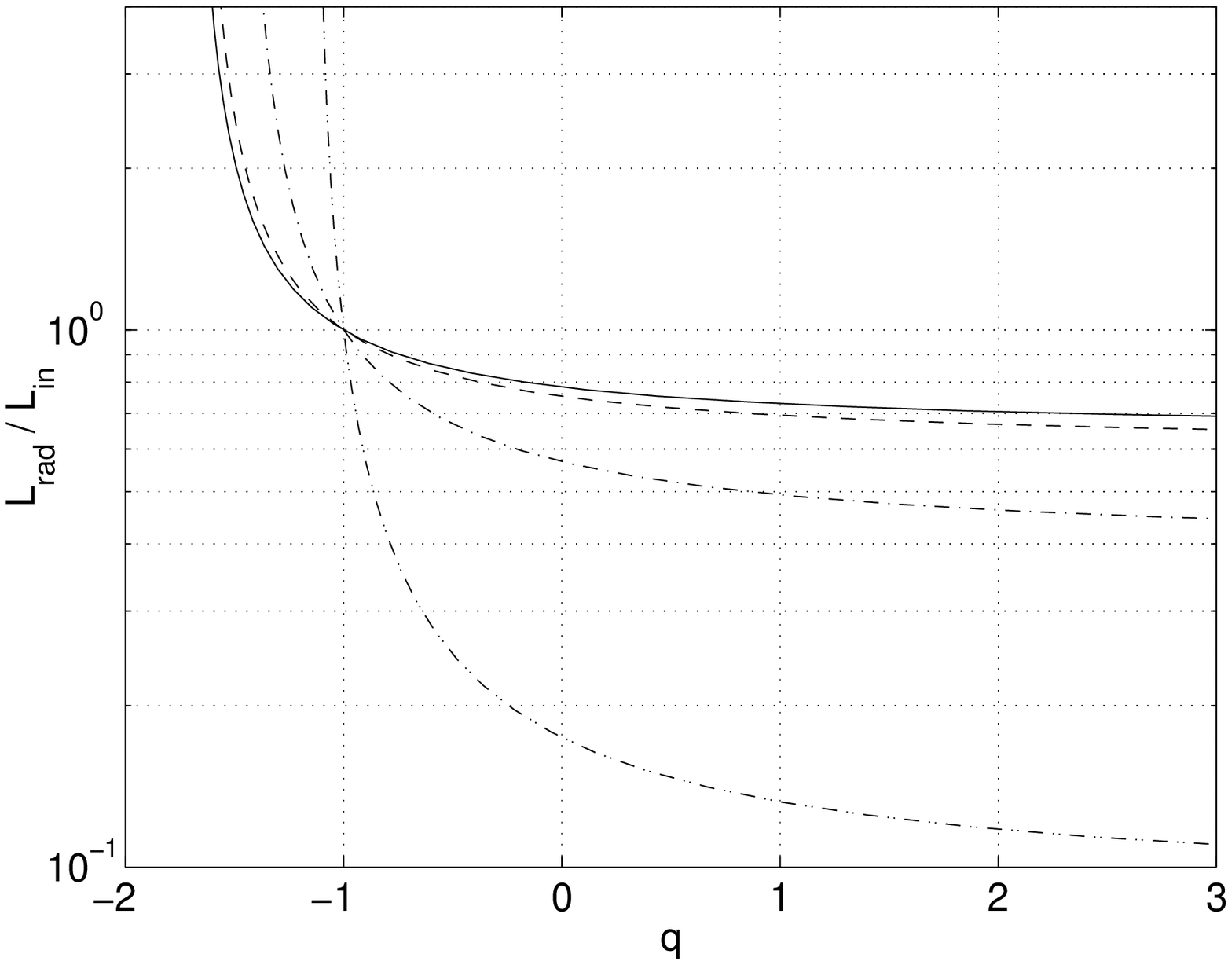}
\caption{ The instantaneous radiative efficiency vs. energy injection rate of 
  an ultra relativistic blast wave, for various semi-radiative
  efficiencies: (i) $\epsilon=0.1$ (dashed double-dotted); (ii)
  $\epsilon=0.5$ ( dashed-dotted); (iii) $\epsilon=0.9$ ( dashed);
  (iv) $\epsilon=1$ (solid).\label{fig:rel_eff} }
\end{center}
\end{figure}

We define the luminosity in a more observable fashion, by comparing
the luminosity measured by an observer to the luminosity that would be
measured if the blast wave shell had not existed, and the source energy
could escape freely to the observer.  Transforming the radiated
luminosity to the observer frame using Eq. \ref{G_of_t}, and dividing
by the source energy injection rate we obtain:
\begin{equation}
\label{rel_eta_tot}
\eta_{tot} = {L_{rad} \over L_{in}}={\chi_s^{1+q} \over 1 + \alpha_{s} (1+q) / \epsilon},
\end{equation}
which is constant in time. Efficiencies for different cases appear in
Fig. \ref{fig:rel_eff}.  Note that the instantaneous efficiencies are
higher than those of Eq. \ref{rel_eta_rad}, which means that energy is
supplied to the shocked matter faster than it is emitted. This
apparent discrepancy is due to time contraction, which causes energy
emitted over a long duration at the source to concentrate into a short
observed time.

\section{Discussion and Conclusions}
\label{sec:conc}
Radiative blast waves with energy injection exhibit self similar
relations both in the Newtonian and in the ultra-relativistic limit.
We have found that if a self similar solution exists, the power-law
index can be one of two: (i) The power-law index of a radiative blast
wave with the same radiative mechanism but without energy injection;
(ii) The power law index of the central source, disregarding radiative
mechanism.  The chosen power law is the one which corresponds to an
evolution with a faster energy increase, or a slower energy decrease.
Sources which emit energy with a steep decreasing power law
($\sigma<-1$), behave according to the same rule.  This corrects a
common belief which states that such sources can be considered as
instantaneous energy release sources. We have shown that this view is
correct for adiabatic blast waves, but that it does not hold for
radiative blast waves.  

The radiative efficiencies in the ultra
relativistic case are higher than those of Newtonian blast waves.
Therefore, the possibility that GRB afterglows are powered by
continuously emitting sources (\cite{KP97a}) cannot by ruled out on
efficiency basis. Moreover, the instantaneous efficiency of a blast
wave with a tuned power law behavior can significantly exceed the
efficiency of the microscopic radiative mechanism.

\begin{appendix}
\section{Analytic solution for the Newtonian fully radiative limit.}
In the fully radiative case the shocked matter is concentrated on a
shell of zero thickness, where the interior contains hot isobaric matter.
In this case the solution can be calculated analytically.

The interior pressure is:
\begin{equation}
\bar P = { (\pgamma -1) \over V(t)} 
\left[ E_0 ({t \over t_0})^{\lambda}  - { 1 \over 2} \rho_1 V(t)  U_{sh}^2 \right]
\end{equation}
where $V(t) = 4 \pi R_{sh}^3 /3 $ is the blast waves volume, and $E_0
({t / t_0})^{\lambda}$ is the energy of the {\it whole} blast wave.
Using the equation of motion 
\begin{equation}
\label{non_dim_motion}
  {d (\rho_1 V U_{sh}) \over dt} = 4 \pi R_{sh}^2 \bar P,
\end{equation}
and Eq. \ref{n_ss} we obtain the radiative efficiency
\begin{equation}
\eta={ L_{rad} \over L_{rad} + \dot E} = { 9 ( \pgamma - 1 ) (
  2+\lambda)^2 \over 2 ( 3 + 4 \lambda) [ 2 ( \lambda-3) + 3 \pgamma (\lambda+2)]}.  
\end{equation}

\section{Analytic solutions for a semi-radiative ultra-relativistic blast wave.}

The self similar ODEs for the  adiabatic interior of a blast wave
 are  (\cite{bm}):
\begin{eqnarray}
\label{r_ss_g}
{\cal G}(y)={1 \over g} {d \ln g \over d \chi} &=&
   { (7m+3k-4)-(m+2)  g \chi  \over (m+1)(4-8  g \chi +  g^2 \chi^2) }
   \nonumber \\
\label{r_ss_f}
{\cal F}(y)={1 \over g} {d \ln f \over d \chi} &=&
  { 8(m-1)+4k-(m+k-4) g \chi \over (m+1)(4-8  g \chi +  g^2 \chi^2) } \\
\label{r_ss_h}
{\cal H}(y)={1 \over g} {d \ln h \over d \chi} &=&
   {2(9m+5k-8)-2(5m+4k-6)  g \chi +  (m+k-2) g^2 \chi^2 \over 
(m+1)(4-8  g \chi +  g^2 \chi^2)  
(2-g \chi)}, \nonumber
\end{eqnarray}
where $y \equiv g(\chi) \chi$. 
These equations are applicable for external density gradient of $\rho \propto r^{-k}$.
Following \cite{bm} we use the relation 
\begin{equation}
g(\chi) {d\chi \over dy}={1 \over y {\cal G}(y) +1}.
\end{equation}
We  obtain
\begin{eqnarray}
{d \ln f \over d  y} &=& { {\cal F}(y) \over y {\cal G}(y) +1}, \label{eq:f}\\
{d \ln h \over d  y} &=& { {\cal H}(y) \over y {\cal G}(y) +1}, \\ 
{d \ln \chi \over d  y} &=& {1  \over y ( y {\cal G}(y) +1)}. \label{eq:chi}
\end{eqnarray}
Using Eq. \ref{eq:chi}, which does not appear in \cite{bm},  we
obtain $\chi(g \chi)$ and a full solution, parameterize by $g
\chi$. As $g \chi$ has physical meaning (Sec. \ref{sec:rel_ss}), this
parameterization is even easier to use than the solution as a
function of the self similar parameter.

We integrate Eq. \ref{eq:f}-\ref{eq:chi} from $y=1+\epsilon$, which
corresponds to the the blast wave edge, toward the contact
discontinuity ($y=2$), and the reverse shock ($y=4$). We obtain
\begin{eqnarray}
{\cal I}_1(y)&=e^ {\int ( d \ln \chi / dy) dy}=& y \gamma^{ (16+28m+m^2 - 3km)/2/\alpha} 
\beta ^ {-m/2-1}, \nonumber \\
{\cal I}_2(y)&=e^{\int ( d \ln f / dy) dy}=& \gamma^{ (64-24m-m^2+k(2m-32+3k))/2/\alpha} 
\beta ^ {(-4-3m+m^2+k(m+1))/2/(m+1)}, \\
{\cal I}_3(y)&=e^{\int ( d \ln h / dy) dy}=& \gamma^{ -h_1/h_2/\alpha}
(y-2)^{ 2(k-m)/h_2}
\beta^{ (24-13m+m^2+k(-19+4m+3k))/h_2}, \nonumber
\end{eqnarray}
using the following definitions:
\begin{equation}
\alpha \equiv ( 160 + 40 m +m^2 - 6 k (12+m) + 9 k^2 )^{1/2},
\end{equation}
\begin{equation}
\beta \equiv -4 -4m+y(12+m-3k) + y^2,
\end{equation}
\begin{equation}
\gamma \equiv { \alpha+ 3k - m - 2 y - 12 \over \alpha - 3k +m +2y +12},
\end{equation}
\begin{equation}
h_1 \equiv 2(480-368m+13m^2+m^3+k(-464+136m+m^2-3k(-41+3m+3k))),
\end{equation}
\begin{equation}
h_2 \equiv 2(-12+3k+m).
\end{equation}
Finally, we obtain the hydrodynamic profiles 
\begin{eqnarray}
\chi( y ) &=& {\cal I}_1(y) /  {\cal I}_1(1+\epsilon) \nonumber \\
f( y ) &=& {\cal I}_2(y) /  {\cal I}_2(1+\epsilon) \nonumber \\
h( y ) &=& {\cal I}_3(y) /  {\cal I}_3(1+\epsilon)  { 1+ \epsilon \over 1 - \epsilon}\\
g( y ) &=& y / \chi(y). \nonumber
\end{eqnarray}

\begin{figure}

\begin{center}
\includegraphics[width=15cm]{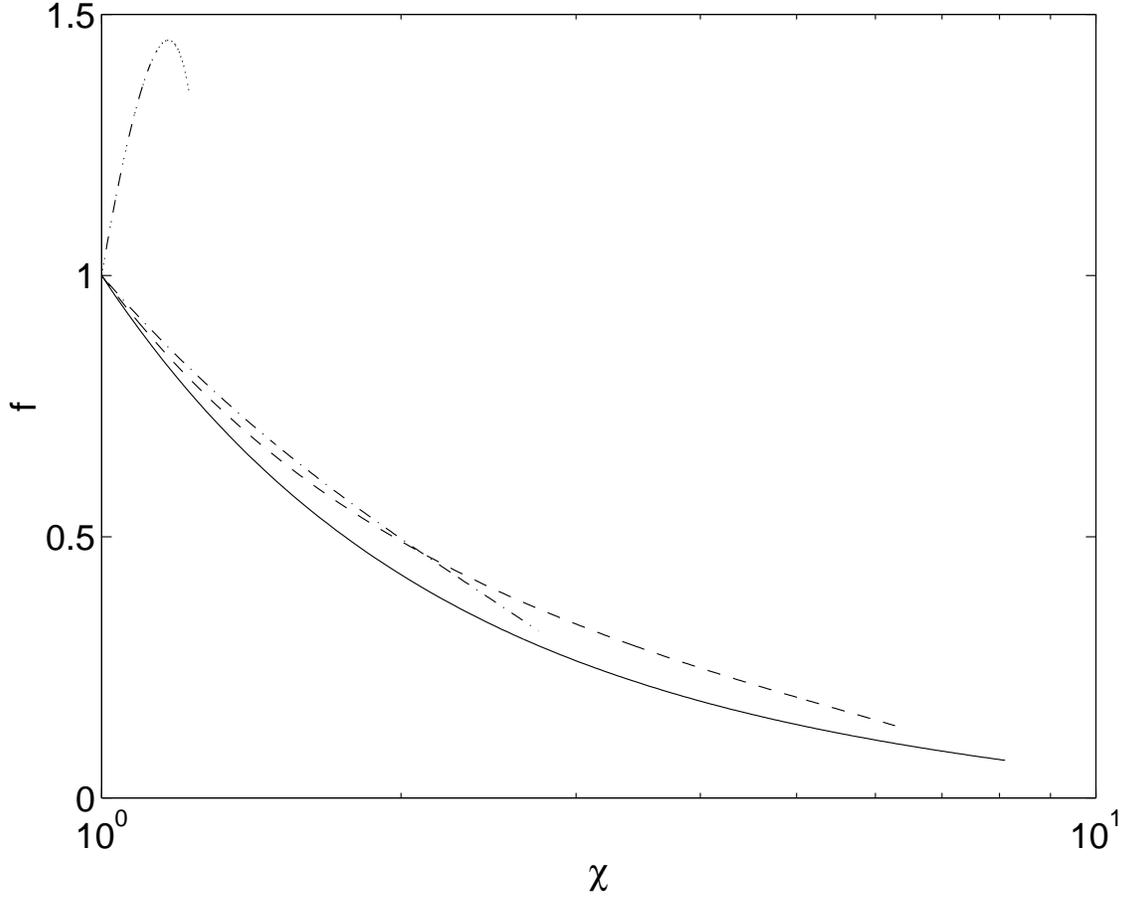}
\caption{
  The normalized pressure $f$ behind the shock front ($\chi=1$) for
  four cases: (i) energy injection $L \propto t^{-6/5}$ semi-radiative
  $\epsilon=0.5$ (solid); (ii) energy injection $L \propto t^{-2/3}$,
  adiabatic (dashed); (iii) energy injection $L \propto t^{-2/3}$,
  semi-radiative $\epsilon=0.5$ (dashed-dotted); (vi) energy injection
  $L \propto t^6$, adiabatic (dashed double-dotted). \label{fig:rel_prof1}}
\end{center}
\end{figure}

\begin{figure}

\begin{center}
\includegraphics[width=15cm]{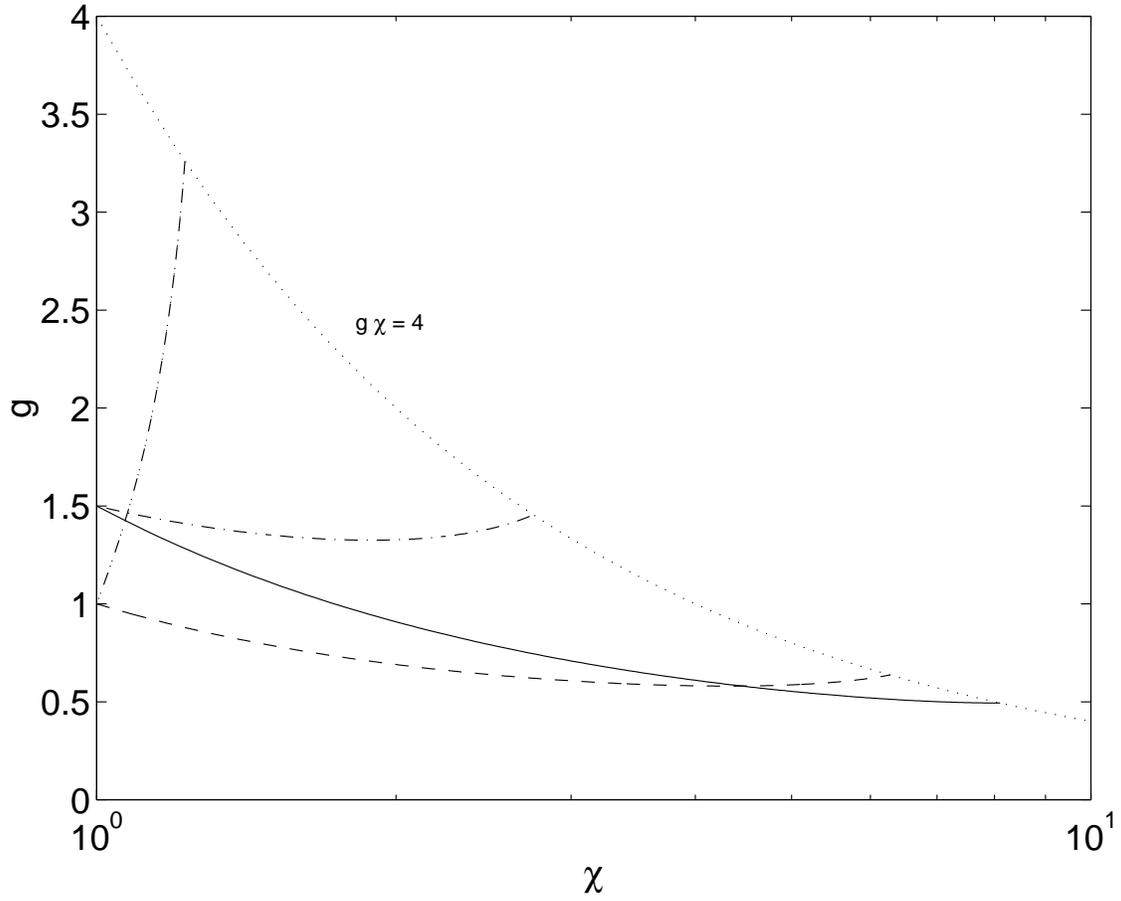}
\caption{
 The normalized Lorenz factor  $g$ behind the shock front, for the
 cases shown in Fig. \ref{fig:rel_prof1}.
 The reverse shock appears at $g(\chi) \chi=4$. \label{fig:rel_prof2}
}.
\end{center}
\end{figure}

\begin{figure}

\begin{center}
\includegraphics[width=15cm]{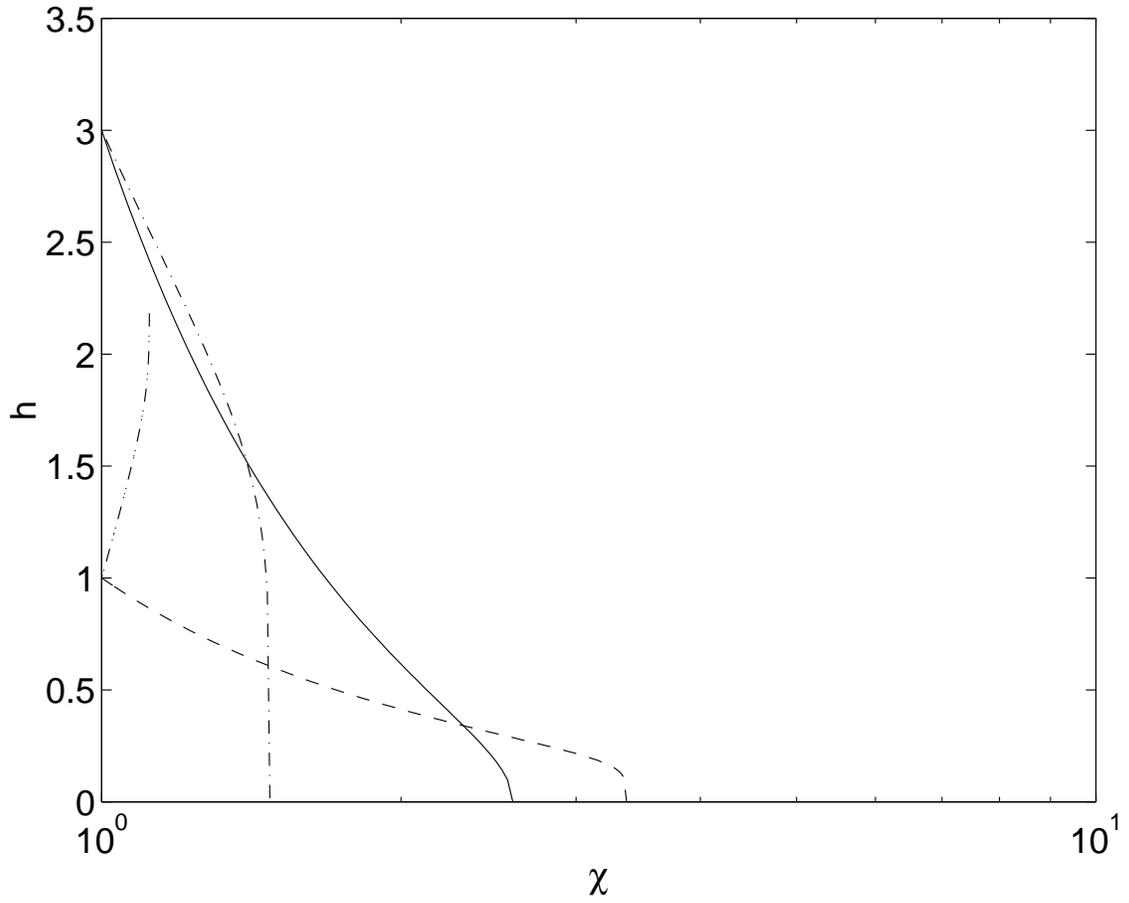}
\caption{
  The normalized density  $h$ behind the shock front, for the
 cases shown in Fig. \ref{fig:rel_prof1}.
 Cases with density that drops to zero correspond to
  $q<2$. At higher injection rates the density diverges to
  infinity at the contact discontinuity.
\label{fig:rel_prof3}
 }
\end{center}
\end{figure}

We have used these relation to calculate the profiles in Fig.
\ref{fig:rel_prof1}-\ref{fig:rel_prof3}.  As discussed in
Sec. \ref{sec:rel_ss}, the parameter $y$ distinguishes
between the different regions within the blast wave.  
For example, the dimensionless energy (Eq. \ref{rel_ener}) held within shocked matter (between $ 1+
\epsilon < y < 4$) is simply
\begin{equation}
\alpha_s=\int_1^{\chi_s} f(\chi) g(\chi) d \chi = \int_{1+\epsilon}^4 {f(y) \over 1+y {\cal G}(y)} dy,
\end{equation}
without the explicit usage of the self similar location of the reverse shock.
\end{appendix}


\newpage

\begin{deluxetable}{lccccc}
\tablecaption{Characteristic relations in continuously powered blast waves.\label{table1}}
\tablehead{
\colhead{condition} &
\colhead{$E_{stored}$ } &
\colhead{$E^0_{rad} / E_{stored}$} &
\colhead{ $E^{\infty}_{rad} /E_{stored} $} & 
\colhead{ $E^0_{inj} / E_{stored}$} &
\colhead{$E^{\infty}_{inj} / E_{stored}$} 
}
\startdata
$\sigma>-1$ & ${L  \over \kappa + \sigma + 1} ({t\over t_L})^{\sigma} t$ & ${\kappa \over 1+\sigma}$
& $\infty$ & $ {\kappa \over \sigma +1 } + 1 > 1 $ & $\infty$ \nl \tablevspace{1ex}
$\sigma=-1$ & ${L  \over \kappa + \sigma + 1} t_L$ & $\infty$ &
$\infty$ & $\infty$ & $\infty$ \nl \tablevspace{1ex}
$-1>\sigma>-1-\kappa$ & ${L  \over \kappa + \sigma + 1} ({t\over
  t_L})^{\sigma} t$ & $\infty$ & ${\kappa \over |1+\sigma|} > 1 $
& $\infty$ & $ {\kappa \over |\sigma +1| } - 1  $ \nl
$\sigma = -1-\kappa$ & \nodata & \nodata & \nodata & \nodata & \nodata
\nl \tablevspace{1ex}
$-1-\kappa>\sigma$ & $E_0 ({t \over t_0})^{-\kappa}$ & $\infty$ & 1 & $\infty$ & 0\\
$L=0$ & $E_0 ({t \over t_0})^{-\kappa}$ & $\infty$ & 1 & 0 & 0 \nl 
\enddata
\end{deluxetable}

\begin{deluxetable}{cccccc}
  \tablecaption{ Parameters of similarity solutions for radiative
    continuously powered Newtonian blast waves with $\pgamma=5/3$. \label{table:newt} }
\tablehead{
\colhead{$\lambda$} &
\colhead{$\sigma$ } &
\colhead{$\epsilon $} & 
\colhead{$1-\xi_{cd}/\xi_0$} &
\colhead{$\eta_{in}$} &
\colhead{$\eta_{tot}$}
}
\startdata  
4    &   3   & 0   & 0.08 & \nodata & \nodata \nl
1    &   0   & 1/2 & 0.05 & 6/11    & 0.34 \nl
1/2  & -1/2  & 0   & 0.20 & \nodata & \nodata \nl
1/2  & -1/2  & 2/5 & 0.10 & 2/3  & 0.23 \nl
-1/4 & -5/4  & 2/5 & 0.62 & 14/9 & 8.35 \nl
-4/7 & -11/7 & 1   & 0    & 1    &  $\infty$ \nl
\enddata
\end{deluxetable}

\begin{deluxetable}{cccccccccc}
  \tablecaption{ Parameters of similarity solutions for radiative
    continuously powered ultra-relativistic blast waves. \label{table:rel} }
\tablehead{
\colhead{m } &
\colhead{q } &
\colhead{ $\epsilon $} & 
\colhead{ $\chi_{cd}$} &
\colhead{$\chi_s$} & 
\colhead{$\alpha_{cd}$} &
\colhead{$\alpha_s$} &
\colhead{K} &
\colhead{$\eta_{rad}$} &
\colhead{$\eta_{tot}$}
}
\startdata
-1 & $\infty$ & 0 & 1 & 1 & 0 & 0 & $\infty$ & 0 & 0  \nl 
2 & -2/3 & 0 & 3.37 & 6.28 & 0.89 & 1.24 & 7.50 & 0 & 0 \nl
2 & -2/3 & 1 & 1    & 1.86 & 0    & 1.20 & 2.22 & 0.71 & 0.88  \nl
3 & -1   & 1 & 1 & 2.38 & 0 & 1.41 & 8 & 1 & 1 \nl
4 & -6/5 & 1/2 & 2.59 & 8.1 & 0.93 & 1.42 & 127.46 & 2.32 & 1.53 \nl
7 & -3/2 & 1 & 1 & 9.04 & 0 & 1.68 & 17417.80 & 6.20 & 2.06\nl
\enddata
\end{deluxetable}




\begin{thebibliography}{}
\bibitem[Begelman \& Cioffi 1989]{BC89}
Begelman, M.C., \& Cioffi, D. F., (1989) \apjl, {\bf 345}, L21
\bibitem[Blandford \& McKee (1976)]{bm}
Blandford, R.D. \& McKee, C.F. (1976) Phys. Fluids,  {\bf 19}, 1130
\bibitem[Bicknell, Dopita \& O'Dea 1997]{BDO97}
Bicknell, G.V., Dopita, M.A., \& O'Dea, C.P.O., (1997) \apj, {\bf 485}, 112
\bibitem[Castor, McCray and Weaver 1975]{C75}
Castor, J., McCray, R., and Weaver, R., (1975), \apjl, {\bf 200}, L107
\bibitem[Cohen, Piran \& Sari 1998]{CPS98}
Cohen, E., Piran, T., \& Sari, R., (1998) \apj, in press.
\bibitem[Katz 1994]{Katz94}Katz, J.I., 1994, ApJ, {\bf 422}, 248.
\bibitem[Katz \& Piran 1997]{KP97a}Katz, J.I., \& Piran, T., 1997, in: C. Meegan, R. Preece \& T. Koshut, Eds.,
{\it Gamma-Ray Bursts 4th Huntsville Symposium} AIP Conf. Proc. 428, pg. 689
(New York: AIP)
\bibitem[Landau \& Lifshitz]{ll}
Landau, L.D. \& Lifshitz E.M. (1987), {\it Fluid Mechanics}, Pergamon press.
\bibitem[McCray 1987]{RM87}
McCray, R., 1987, in {\it Supernova remnants and the interstellar
  medium}, ed. Roger, R.S., and Landecker, T.L., (Cambrige: Cambridge
university press), p. 447
\bibitem[M\'esz\'aros \& Rees 1997]{MR97}M\'esz\'aros, P., \& Rees, M.J., 1997, \apj, {\bf 476}, 232.
\bibitem[M\'esz\'aros, Rees, \& Wijers 1997]{mes}
M\'esz\'aros, P., Rees, M. J., \& Wijers, A. M. J.,
(1997),  \apj, {\bf 499}, 301.
\bibitem[Rees \& M\'esz\'aros (1998)]{RM98}
Rees, M.J. \& M\'esz\'aros, P. (1998) \apj, {\bf 496}, L1
\bibitem[Ostriker \& McKee (1988)]{OM88}
Ostriker, J. P., \& McKee C.F., (1988) Rev. Mod. Phys., {\bf  60}, 1
\bibitem[Paczy\'nski \& Rhoads 1993]{PacRho93}Paczy\'nski, B., \& Rhoads, J., 1993,
\apjl, {\bf 418}, L5.
\bibitem[Panaitescu, M\'esz\'aros and Rees (1998)]{PMR98}
Panaitescu, A., M\'esz\'aros, P., and Rees, M.J., astro-ph/9801258
\bibitem[Piran 1998]{piran98}
Piran, T., (1998)  Physics Reports, in press.
\bibitem[Sari, Piran, \& Narayan 1998]{sari}
Sari, R., Piran, T., \& Narayan, R., (1998), \apjl, {\bf 497}, L17.
\bibitem[Scheuer 1974]{S74}
Scheuer, P. A. G., (1974) M.N.R.A.S. {\bf 166},513
\bibitem[Waxman 1997]{waxman}
Waxman, E., 1997, \apj, {\bf 485}, L9.
\bibitem[Weaver \etal 1977]{W77}
Weaver, R., McCray, R., Catsor, J., Shapiro, P., and Moore, R. (1977)
\apj, {\bf 218} 377
\bibitem[Weiler 1983]{Wei83}
Weiler, K. W., 1983, in IAU Symposium 101,{\it Supernova Remnants and their X-Ray
  Emission}, ed. Danziger, J., and Gorenstein, P., (Dordrecht:Reidel),
pg. 299
\bibitem[Wijers, Rees and M\'esz\'aros 1997]{Wiejers_MR97}
Wijers, A.M.J., Rees, M.J., \& M\'esz\'aros, P., 1997, MNRAS, {\bf 288}, L51.
\bibitem[in 't Zand \etal 1998]{Zand98}
in 't Zand, J.J.M., (1998), astro-ph/9807314
\end{thebibliography}
\end{document}